\documentclass[a4paper,aps,showpacs,floatfix,pre,twocolumn]{revtex4}

\usepackage{graphicx}

\begin{document}

\title{Simple and Robust Solver for the Poisson--Boltzmann Equation}

\author{M. Baptista, R. Schmitz, B. D\"unweg}
\affiliation{Max Planck Institute for Polymer Research \\
Ackermannweg 10, 55128 Mainz, Germany}

\date{\today}

\begin{abstract}
  A variational approach is used to develop a robust numerical
  procedure for solving the nonlinear Poisson--Boltzmann
  equation. Following Maggs et al., we construct an appropriate
  constrained free energy functional, such that its Euler--Lagrange
  equations are equivalent to the Poisson--Boltzmann equation. This is
  a formulation that searches for a true minimum in function space, in
  contrast to previous variational approaches that rather searched for
  a saddle point. We then develop, implement, and test an algorithm
  for its numerical minimization, which is quite simple and
  unconditionally stable. The analytic solution for planar geometry is
  used for validation. Some results are presented for a charged
  colloidal sphere surrounded by counterions, and optimizations based
  upon Fast Fourier Transforms and hierarchical pre--conditioning are
  briefly discussed.
\end{abstract}


\maketitle

\section{Introduction}

In many soft-matter and biological systems electrostatic interactions
have a strong influence on the physical behavior. A typical example
are charge--stabilized colloidal dispersions \cite{Russel1989}, whose
structure is mainly determined by the interplay between van der Waals
attraction and electrostatic repulsion. A simplifying feature is the
fact that it is often sufficient to describe the structure of the
``cloud'' of surrounding ions (counterions and salt ions) just in
terms of Mean Field theory, in particular if the ions are
monovalent. At the center of this theory is the well--known
Poisson--Boltzmann equation, which, from the mathematical point of
view, is a nonlinear partial differential equation. Analytical work in
this field has been mainly confined to the linearized (Debye--H\"uckel)
version, which however is often insufficient, in particular near strongly
charged objects. This has prompted efforts to develop methods to solve
the equation numerically, ideally in three dimensions and without
restrictions on the underlying geometry or spatial symmetry. These are
usually based on standard finite--difference \cite{rocchia} or
finite--element \cite{holst} techniques (for a recent review see
Ref. \cite{Lu}), and have meanwhile reached a substantial degree of
sophistication and complexity.

Recently, however, Maggs \cite{Maggs2002} has put forward a completely
new approach to electrostatics in soft--matter research, which bears a
certain similarity to lattice gauge theories \cite{duncan}. The
central idea is to use the \emph{electric field} $\mathbf{E}$ instead
of the electrostatic potential $\psi$ as the quantity on which the
algorithm operates, and to view Gauss' law $\varepsilon \nabla \cdot
\mathbf{E} = \rho$ as a \emph{constraint} for the field
configurations. On a lattice, it is then easy to construct an electric
field ($\mathbf{E}$) configuration which satisfies Gauss' law for a
given (arbitrary) charge density $\rho$. The hard part is rather the
transversal part of the field, which should satisfy $\nabla \times
\mathbf{E} = 0$, but initially does not (unless one uses a
sophisticated initialization procedure based upon solving the Poisson
equation). This transversal degree of freedom can then be removed by
local relaxations (this is the approach taken in the present paper),
or integrated out by performing a Monte Carlo \cite{Maggs2002,maggs_jcp}
or Molecular Dynamics \cite{rottler2004lmd,rottler2004cbs,Pasichnyk2004}
simulation on the overall system. A crucial aspect of the method is to
locally update both $\rho$ and $\mathbf{E}$ simultaneously in a way that
Gauss' law is still satisfied \emph{after} the update, i.~e. the
``constraint surface'' is never left.

The original Maggs approach is based upon a system of discrete charges
whose statistical physics is treated in a consistent way. However, it
is also possible to apply this to the Mean Field version of the
theory, i.~e. the Poisson--Boltzmann equation. The purpose of the
present paper is to outline how this can be done in practice. In
Sec. \ref{sec:derivation} we derive the method by re--formulating the
Poisson--Boltzmann theory in terms of a free energy functional, which
is minimized by using a Maggs--type algorithm. In contrast to previous
formulations, this functional provides a true minimum, and therefore
the procedure is very straightforward and simple, unconditionally
stable, and has a rather modest storage requirement which scales only
linearly with the number of grid points. In terms of computational
speed, the method can probably not yet compete with the existing
packages; however, it is reasonable to assume that more advanced
versions that combine the basic methodology with acceleration
techniques like adaptive mesh refinement and / or unstructured meshes,
may become a very useful tool. Section \ref{sec:numerics} presents
numerical results obtained with our current simple implementation,
while Sec. \ref{sec:speedup} discusses optimizations based upon Fast
Fourier Transforms and hierarchical pre--conditioners. Finally,
Sec. \ref{sec:conclus} finishes with some concluding remarks.

\section{Derivation of the Algorithm}
\label{sec:derivation}

\subsection{Poisson--Boltzmann Equation}

Consider a system of fixed charges, with total charge $Z e$ ($e > 0$
denotes the elementary charge) dispersed in a solvent with dielectric
constant $\varepsilon$. The system is confined to a three--dimensional
finite domain of volume $V$. The charges can be distributed either in
macroscopic particles or in any other kind of boundary.  Furthermore,
the domain contains counterions of total charge $- Z e$ such that the
system as a whole is charge neutral. The total number of counterions
is $N_{0}$, such that $Z = - z_{0} N_{0}$, where $z_{0}$ is their
valence. Furthermore, the presence of other ionic species (salt ions)
is allowed if they satisfy charge neutrality, i.~e. if $\sum_{\alpha
  \geq 1} z_{\alpha} N_{\alpha} = 0$, where $z_{\alpha}$ is the
valence of the ionic species $\alpha$ and $N_{\alpha}$ is its number.

The continuum theory provides equations for the concentrations
(particle number densities) $c_{\alpha}(\mathbf{r})$ of each ionic
species $\alpha$ and the electrostatic potential
$\psi(\mathbf{r})$. In the stationary regime (no time dependence) and
in thermal equilibrium characterized by the thermal energy $k_B T$,
they take the form \cite{Russel1989}
\begin{eqnarray}
  k_{B} T \nabla \ln c_{\alpha} + e z_{\alpha} \nabla \psi & = & 0 ,
  \label{eq:nernst_planck} \\
  \varepsilon \nabla^{2} \psi + \sum_{\alpha} e z_{\alpha} c_{\alpha} 
  & = & 0 . \label{eq:poisson}
\end{eqnarray}
The first equation is the equilibrium version of the Nernst--Planck
equation, which balances the diffusion current of ionic species
$\alpha$ against the drift caused by the electric field $\mathbf{E} =
- \nabla \psi$. The second equation is the Poisson equation, taking
into account all ionic species as a source term, while the fixed
charges appear as boundary conditions. The total number of particles
is obtained by integrating the concentration over the whole domain:
\begin{equation}
  N_{\alpha} = \int_{V} c_{\alpha}dV,
\end{equation}
and the charge neutrality may then be expressed by
\begin{equation}
  \sum_{\alpha} z_{\alpha} \int_{V} c_{\alpha} dV = -Z,
\end{equation}
with
\begin{equation}
  \sum_{\alpha\geq1} z_{\alpha} \int_{V} c_{\alpha} dV = 0 .
\end{equation}
By integrating Eq. \ref{eq:nernst_planck} one obtains
\begin{equation}
  c_{\alpha} = A_{\alpha} \exp \left( - \frac{e z_{\alpha} \psi}{k_{B}T} \right) ,
\end{equation}
where the integration constant $A_{\alpha}$ has the dimension of a
concentration, and, for normalization reasons, must have the
value
\begin{equation}
  A_{\alpha} = \frac{N_{\alpha}}{
  \int_{V} \exp \left( - e z_{\alpha} \psi / k_{B}T \right) dV} .
\end{equation}
Inserting this result into Eq. \ref{eq:poisson}, one obtains the
Poisson--Boltzmann equation:
\begin{equation} \label{eq:compact_poisson_boltzmann}
  \varepsilon \nabla^{2} \psi + \sum_{\alpha} e z_{\alpha} A_{\alpha}
  \exp \left( - \frac{e z_{\alpha} \psi}{k_{B}T} \right) = 0,
  \label{eq:poisson_boltzmann}
\end{equation}
which is an equation for $\psi$ only. However, for the algorithm to be
discussed below, it will be advantageous to rather consider the
equivalent coupled set of equations (Eqs. \ref{eq:nernst_planck},
\ref{eq:poisson}).

\subsection{Reduced Units}

The Poisson--Boltzmann equations can be rewritten in terms of
nondimensional quantities, i.~e. in reduced units. The argument of the
exponential in Eq. \ref{eq:poisson_boltzmann} suggests the most
natural way of rescaling the potential:
\begin{equation}
  \psi' = \frac{e \psi}{k_{B} T} ,
\end{equation}
i.~e. the reduced potential is the electrostatic energy of an
elementary charge in units of the thermal energy. This leads
to
\begin{eqnarray}
  \label{eq:nernstplanck1}
  \nabla \ln c_{\alpha} + z_{\alpha} \nabla \psi' & = & 0 , \\
  \label{eq:poisson1}
  \nabla^{2} \psi' + 4 \pi l_{B} \sum_{\alpha} z_{\alpha} c_{\alpha} & = & 0 ,
\end{eqnarray}
where $l_{B} = e^{2} / (4 \pi \varepsilon k_{B}T)$ is the Bjerrum
length. Introducing a parameter $\kappa^{-1}$ as a characteristic
length scale (see below), the gradient operator is rescaled via
$\nabla = \kappa \nabla'$. This allows writing the equations
in the nondimensional form
\begin{eqnarray}
  \label{eq:nernstplanck2}
  \nabla' \ln c'_{\alpha} + z_{\alpha} \nabla' \psi' & = & 0 , \\
  \label{eq:poisson2}
  \nabla'^{2} \psi' + \sum_{\alpha} z_{\alpha} c'_{\alpha} & = & 0 ,
\end{eqnarray}
where
\begin{equation}
  c_{\alpha}' = 4 \pi l_{B} \kappa^{-2} c_{\alpha} 
\end{equation}
is the reduced concentration. In terms of the electric
field $\mathbf{E}' = - \nabla' \psi'$, the equations
are
\begin{eqnarray}
  \label{eq:nernstplanck3}
  \nabla' \ln c'_{\alpha} & = & z_{\alpha} \mathbf{E}' ,\\
  \label{eq:gauss3}
  \nabla' \cdot \mathbf{E}' & = & \sum_{\alpha} z_{\alpha} c'_{\alpha} , \\
  \label{eq:curl3}
  \nabla' \times \mathbf{E}' & = & 0 .
\end{eqnarray}
The normalization condition for the amount of species $\alpha$ is
transformed to
\begin{equation}
  \int_{V'} c'_{\alpha} dV' = N'_{\alpha}
\end{equation}
with
\begin{equation}
  N'_{\alpha} = 4 \pi l_B \kappa N_{\alpha} .
\end{equation}

The choice of the parameter $\kappa$ is completely immaterial for the
mathematical formulation of the problem. It is only important to map
the numerical results back onto a physical system, and therefore a
matter of convention. For many applications, the choice
\begin{equation} \label{eq:definekappa} 
  \kappa^{2} = 4 \pi l_B \frac{\sum_{\alpha} z_{\alpha}^2 N_{\alpha}}{V}
\end{equation}
turns out to be quite useful: This is the natural screening parameter
in the finite--volume version of linearized Poisson--Boltzmann theory
(therefore it also appears in the corresponding treatment of
asymmetric electrolytes \cite{beresford}). In the case of only one
monovalent ionic species (the counterions $\alpha = 0$), this reduces
to
\begin{equation}
  \kappa^{2} = 4 \pi l_{B} \frac{\left\vert Z \right\vert}{V} .
\end{equation}
It should be noted that in this case $N'_0 = V'$.

From now on, we will be concerned with the problem of numerically
solving the reduced set
Eqs. \ref{eq:nernstplanck3}--\ref{eq:curl3}. In what follows, the
primes will be omitted, with the understanding that all quantities
(including $N_{\alpha}$ and $V$) are given in reduced units.

\subsection{Variational Approach}
\label{sec:variational_approach}

Following the ideas of Maggs and Rosetto \cite{Maggs2002}, the
Poisson--Boltzmann equation can be re--formulated as a constrained
variational problem, where a free energy functional is minimized.
This functional is constructed such that its Euler--Lagrange equations
are equivalent to Eqs. \ref{eq:nernstplanck3}--\ref{eq:curl3}. Its
form is
\begin{eqnarray}
  \label{eq:define_functional_1}
  F & = & \int_{V} \, f\, dV , \\
  \nonumber
  \label{eq:define_functional_2}
  f & = & \frac{1}{2} \mathbf{E}^{2}
  + \sum_{\alpha} c_{\alpha} \ln c_{\alpha}
  - \psi \left(\nabla \cdot \mathbf{E} 
        - \sum_{\alpha} z_{\alpha} c_{\alpha} \right) \\
  & - & \sum_{\alpha} \mu_{\alpha}
        \left( c_{\alpha} - \frac{N_{\alpha}}{V} \right) .
\end{eqnarray}
The first term corresponds to the electrostatic energy and the second
to the entropy. Gauss' law and the mass normalization conditions are
included as constraints, via Lagrange multipliers: The field $\psi
\left( \mathbf{r} \right)$ is the electrostatic potential, while the
numbers $\mu_{\alpha}$ are the chemical potentials of the species
$\alpha$.

The Euler--Lagrange equations for this variational problem will be
derived next. Variation with respect to $\psi$ and $\mu_{\alpha}$ just
recovers the constraint equations
\begin{eqnarray}
  \label{eq:constraint1}
  \nabla \cdot \mathbf{E} & = & \sum_{\alpha} z_{\alpha} c_{\alpha} , \\
  \label{constraint2}
  \int_{V} c_{\alpha} dV & = & N_{\alpha} .
\end{eqnarray}
Variation with respect to $c_{\alpha}$ results in
\begin{equation} \label{eq:variation1}
  \ln c_{\alpha} + 1 + \psi z_{\alpha} - \mu_{\alpha} = 0 ,
\end{equation}
while variation with respect to $\mathbf{E}$ yields
\begin{equation} \label{eq:variation2}
  \mathbf{E} = - \nabla \psi .
\end{equation}
The dependence on the unknown Lagrange multipliers is removed by
taking the gradient of Eq. \ref{eq:variation1} and the curl of
Eq. \ref{eq:variation2} to obtain
\begin{eqnarray}
  \label{eq:variation3}
  \nabla \ln c_{\alpha} + z_{\alpha} \nabla \psi & = & 0 , \\
  \label{eq:variation4}
  \nabla \times \mathbf{E} & = & 0. 
\end{eqnarray}
Finally, inserting Eq. \ref{eq:variation2} into Eq.
\ref{eq:variation3} leads to
\begin{equation}
  \label{eq:variation5}
  \nabla \ln c_{\alpha} = z_{\alpha} \mathbf{E} .
\end{equation}
In summary, the derived equations (Eqs. \ref{eq:variation5},
\ref{eq:constraint1}, \ref{eq:variation4}) are the desired set --- in
other words, the solution of the minimization problem is identical to
the solution of the Poisson--Boltzmann equation.

In this context, one should make a few important observations.
Firstly, we note that solving the Euler--Lagrange equations will
provide one and \emph{only one} solution: It has been proven (by
variational methods) that the Poisson--Boltzmann equation has one
unique solution \cite{holstpdf,duan_exist}. Secondly, it is easy to
show that the solution is indeed a local minimum: By writing
\begin{eqnarray}
\mathbf{E} & = & \mathbf{E}^{(0)} + \delta \mathbf{E} , \\
c_{\alpha} & = & c_{\alpha}^{(0)} + \delta c_{\alpha} ,
\end{eqnarray}
where $\mathbf{E}^{(0)}$, $c_{\alpha}^{(0)}$ form the solution
of the problem, and $\delta \mathbf{E}$, $\delta c_{\alpha}$
are small deviations which satisfy the constraints, i.~e.
\begin{eqnarray}
\nabla \cdot \delta \mathbf{E} & = &
\sum_{\alpha} z_{\alpha} \delta c_{\alpha} , \\
\int_V \delta c_{\alpha} dV & = & 0 ,
\end{eqnarray}
one finds
\begin{eqnarray}
\nonumber
F & = & \int_V \left\{ \frac{1}{2} \mathbf{E}^{(0)2}
         + \sum_{\alpha} c_{\alpha}^{(0)} \ln c_{\alpha}^{(0)} \right\}
         \, dV \\
  & + & \int_V \left\{ \frac{1}{2} \delta \mathbf{E}^2
        + \frac{1}{2} \sum_{\alpha} 
          \frac{\delta c_{\alpha}^2}{c_{\alpha}^{(0)}} \right\} \, dV
        + O \left( \delta c^3 \right) ,
\end{eqnarray}
i.~e. small deviations from the solution will always increase the
functional. Together with the uniqueness, this shows that the
functional has one and only one minimum, which corresponds to the
solution of the Poisson--Boltzmann problem. Therefore, a procedure
that relaxes all degrees of freedom of the functional such that it is
systematically decreased, while staying on the constraint surface,
will ultimately run into the one and only minimum of the free energy
landscape. The iterative procedure may be slow and hampered by small
eigenvalues of the Hessian at the minimum, but such problems can be
kept under control by careful convergence checks and variation of the
number of iterations. A simple algorithm that initializes the system
on the constraint surface, keeps it there, and systematically
decreases $F$ by local updates of all non--constrained degrees of
freedom will be outlined in Sec. \ref{sub:algorithm}. Since the
constraints are always satisfied during the procedure, the
Lagrange--multiplier terms in the functional may be omitted, such that
it is simplified to
\begin{equation}
  F = \int_{V} \left\{
  \frac{1}{2} \mathbf{E}^{2} + \sum_{\alpha} c_{\alpha} \ln c_{\alpha} 
  \right\} \, dV .
\end{equation}

It should also be noted that previous approaches that were also based
upon a free--energy functional (see, e.~g., Ref.
\cite{che_potential_free_energy}), did \emph{not} truly search for a
minimum, but rather for a \emph{saddle point}. These methods were not
based upon a functional that involves the electric field, but rather
one that depends on the electrostatic potential,
\begin{eqnarray}
F & = & \int_V f \, dV \\
\nonumber
f & = & - \frac{1}{2} \left( \nabla \psi \right)^2
        + \sum_{\alpha} c_{\alpha} \ln c_{\alpha}
        + \psi \sum_{\alpha} z_{\alpha} c_{\alpha} \\
  & - & \sum_{\alpha} \mu_{\alpha}
        \left( c_{\alpha} - \frac{N_{\alpha}}{V} \right) ,
\end{eqnarray}
where the $\mu_{\alpha}$ are, as before, the Lagrange multipliers
corresponding to the mass normalization conditions. However, in
contrast to its meaning in Eqs. \ref{eq:define_functional_1},
\ref{eq:define_functional_2}, $\psi$ is here \emph{not} a Lagrange
multiplier, but rather a degree of freedom. It is straightforward to
show that the Euler--Lagrange equations of this problem are equivalent
to the Poisson--Boltzmann equation. To show that this solution is
indeed not a minimum but rather a saddle point, one again decomposes
$\psi$ and $c_{\alpha}$ into the solution plus small deviations,
\begin{eqnarray}
\psi & = & \psi^{(0)} + \delta \psi , \\
c_{\alpha} & = & c_{\alpha}^{(0)} + \delta c_{\alpha} ,
\end{eqnarray}
with
\begin{equation}
\int_V \delta c_{\alpha} dV = 0 .
\end{equation}
This yields
\begin{eqnarray}
F   & = & \int_V \left( f_0 + f_2 \right) dV 
          + O \left( \delta c^3 \right) , \\
f_0 & = & 
        - \frac{1}{2} \left( \nabla \psi^{(0)} \right)^2
        + \sum_{\alpha} c_{\alpha}^{(0)} \ln c_{\alpha}^{(0)} \\
\nonumber
    & + & \psi^{(0)} \sum_{\alpha} z_{\alpha} c_{\alpha}^{(0)} , \\
f_2 & = & 
        - \frac{1}{2} \left( \nabla \delta \psi \right)^2
        + \frac{1}{2} \sum_{\alpha} 
          \frac{\delta c_{\alpha}^2}{c_{\alpha}^{(0)}} \\
\nonumber
    & + & \delta \psi \sum_{\alpha} z_{\alpha} \delta c_{\alpha} ,
\end{eqnarray}
i.~e. the quadratic form of the deviations is \emph{not}
positive--definite. It is natural to suspect that this lack of
positive--definiteness causes various numerical difficulties in terms
of stability, which therefore are intrinsically absent in the new
formulation.

\subsection{Discretization}

The computational domain is a rectangular parallelepiped of size
$l_{1} \times l_{2} \times l_{3}$ with periodic boundary
conditions. This box is discretized by a simple orthorombic (usually:
cubic) lattice with sites $\mathbf{r}_0$ and lattice spacings $\Delta
x_{i}$, $i = 1, 2, 3$ enumerating the Cartesian directions. The volume
of a unit cell is thus $\Delta V = \Delta x_1 \, \Delta x_2 \, \Delta
x_3$. The concentrations $c_{\alpha}$ are variables on the sites,
while the electric field is associated with the \emph{links}.
The positions of the concentration fields are the vectors
\begin{equation}
  \mathbf{r}_0(\mathbf{n}) = \left( \Delta x_1 n_1,
                                    \Delta x_2 n_2,
                                    \Delta x_3 n_3 \right),
\end{equation}
where $n_i$ are integers. The field $E_1$ is located at the positions
\begin{equation}
  \mathbf{r}_1(\mathbf{n}) = \left( \Delta x_1 (n_1 + 1/2), 
                                    \Delta x_2 n_2, 
                                    \Delta x_3 n_3 \right). 
\end{equation}
Similarly, the positions for $E_2$ and $E_3$ are
\begin{eqnarray}
  \mathbf{r}_2(\mathbf{n}) & = & \left( \Delta x_1 n_1, 
                                        \Delta x_2 (n_2 + 1/2),
                                        \Delta x_3 n_3 \right), \\
  \mathbf{r}_3(\mathbf{n}) & = & \left( \Delta x_1 n_1, 
                                        \Delta x_2 n_2, 
                                        \Delta x_3 (n_3 + 1/2) \right) ,
\end{eqnarray}
respectively. Furthermore, it is useful to define
\begin{eqnarray}
  \mathbf{r}'_1(\mathbf{n}) & = & \left( \Delta x_1 (n_1 - 1/2), 
                                         \Delta x_2 n_2,
                                         \Delta x_3 n_3 \right), \\
  \mathbf{r}'_2(\mathbf{n}) & = & \left( \Delta x_1 n_1, 
                                         \Delta x_2 (n_2 - 1/2),
                                         \Delta x_3 n_3 \right), \\
  \mathbf{r}'_3(\mathbf{n}) & = & \left( \Delta x_1 n_1, 
                                         \Delta x_2 n_2, 
                                         \Delta x_3 (n_3 - 1/2) \right) .
\end{eqnarray}
These definitions allow to approximate the functional by
\begin{eqnarray}
  \frac{F}{\Delta V} & = &
  \frac{1}{2} \sum_{\mathbf{n}} \sum_{i=1}^3
  E_i^2 \left( \mathbf{r}_i(\mathbf{n}) \right) \\
  \nonumber
  & + &
  \sum_\alpha \sum_{\mathbf{n}}
  c_\alpha \left( \mathbf{r}_0(\mathbf{n}) \right) \ln
  c_\alpha \left( \mathbf{r}_0(\mathbf{n}) \right) ,
\end{eqnarray}
and to also discretize the divergence operator in a straightforward
way:
\begin{equation}
  \left( \nabla \cdot \mathbf{E} \right)
  \left( \mathbf{r}_0(\mathbf{n}) \right) =
  \sum_{i=1}^3 \frac{1}{\Delta x_i} \left( 
  E_i \left( \mathbf{r}_i(\mathbf{n}) \right) -
  E_i \left( \mathbf{r}'_i(\mathbf{n}) \right) \right) .
\end{equation}
Gauss' law then reads
\begin{equation}
  \sum_{i=1}^3 \frac{1}{\Delta x_i} \left( 
  E_i \left( \mathbf{r}_i(\mathbf{n}) \right) -
  E_i \left( \mathbf{r}'_i(\mathbf{n}) \right) \right) =
  \sum_\alpha z_\alpha c_\alpha \left( \mathbf{r}_0 (\mathbf{n}) \right) .
\end{equation}
Introducing fluxes via
\begin{eqnarray}
\phi_1 & = & E_1 \Delta x_2 \Delta x_3 , \\
\phi_2 & = & E_2 \Delta x_3 \Delta x_1 , \\
\phi_3 & = & E_3 \Delta x_1 \Delta x_2 ,
\end{eqnarray}
this is rewritten as
\begin{equation}
  \sum_{i=1}^3 \left( 
  \phi_i \left( \mathbf{r}_i(\mathbf{n}) \right) -
  \phi_i \left( \mathbf{r}'_i(\mathbf{n}) \right) \right) =
  \Delta V \, \sum_\alpha z_\alpha c_\alpha 
  \left( \mathbf{r}_0 (\mathbf{n}) \right) .
\end{equation}
Finally, the normalization condition for the amount of ionic species
$\alpha$ is discretized as
\begin{equation}
  \sum_{\mathbf{n}} 
  c_\alpha \left( \mathbf{r}_0(\mathbf{n}) \right) =
  \frac{N_\alpha}{\Delta V} .
\end{equation}

\subsection{Algorithm}
\label{sub:algorithm}

The numerical minimization procedure starts from some configuration of
the discretized fields $c_\alpha$ and $\mathbf{E}$ which satisfies all
the constraints, i.~e. the normalization conditions for the ions, plus
Gauss' law. The algorithm then performs successive local changes in
the electric fields and the concentrations, analogously to the Monte
Carlo moves of Maggs and Rosetto \cite{Maggs2002}. These moves have
the big advantage that they rigorously conserve the constraints. In
contrast to Ref. \cite{Maggs2002}, however, the moves are not
stochastic, but rather deterministic, and constructed in such a way
that they decrease the functional, and do this optimally. Since the
free energy landscape of this problem has a simple structure (as
discussed in Sec. \ref{sec:variational_approach}), the procedure
relaxes the fields into the one and only minimum, which is the
solution of the Poisson--Boltzmann equation.

The algorithm can be summarized as follows:
\begin{enumerate}
\item Distribute the fixed charges.
\item Classify the grid points.
\item Distribute the ionic species uniformly in the moveable nodes.
\item Initialize the electric field.
\item Perform the field moves for all the plaquettes (smallest
      closed loops) in the grid.
\item Perform the concentration moves for all pairs of adjacent
      moveable nodes.
\item Check if the changes in the functional caused by steps 5 and
      6 are less than a given tolerance: if yes, then stop, otherwise,
      return to step 5.
\end{enumerate}

In the beginning, fixed charges and ionic species must be distributed
over the grid. Fixed charges are usually associated with surfaces.
Therefore elements of surface charge density must be mapped onto
elements of volume charge density, so that they can be associated with
some of the nodes. These nodes are then marked as ``fixed'' and no
particles can enter or leave them after initialization. Further nodes
may be marked as ``fixed'' if they are known to be empty (for example,
if they represent the interior of a particle). The nodes representing
the volume where ions can move are marked as ``moveable''. Initially,
each ionic species is uniformly distributed over the ``moveable''
nodes. Choosing the correct amount of charges then automatically
results in charge neutrality of the overall system.

The next step consists in initializing the electric field so that it
satisfies Gauss' law for the initial charge distribution. One
possibility is to solve the Poisson equation with some numerical
method. This needs to be done only once, in the initialization. An
alternative, based on charge neutrality, is to initialize each
component by means of a recursion over the spatial dimensions
\cite{Pasichnyk2004}, which is equivalent to applying Gauss' law to
linear chains of nodes. First the lattice is decomposed into a set of
planes perpendicular to the $x_1$--axis and it is required that $E_1$
takes the same identical value for all links with identical
$x_1$--coordinate. Then in one (arbitrary) plane of links we set $E_1
= 0$. Starting from there, we can then calculate $E_1$ step by step in
the subsequent planes of links, where the change in $E_1$ is given by
the plane--averaged charge density between the links. Assuming that
the procedure is started at the charge plane $x_1 = 0$, it reads
\begin{eqnarray}
  &&
  E_1 (- 0.5 \Delta x_1, n_2 \Delta x_2, n_3 \Delta x_3) = 0 , \\
  \nonumber
  &&
  E_1 ( (n_1 + 0.5) \Delta x_1, n_2 \Delta x_2, n_3 \Delta x_3)
  = \\
  \nonumber
  &&
  E_1 ( (n_1 - 0.5) \Delta x_1, n_2 \Delta x_2, n_3 \Delta x_3)
  + \\
  &&
  \Delta x_1 \, \left< \rho \right> (n_1 \Delta x_1) ,
\end{eqnarray}
where $\left< \rho \right>$ is the plane--averaged charge density at
$x_1 = n_1 \Delta x_1$. Charge neutrality combined with the periodic
boundary conditions ensures that this procedure will give consistent
results after closing the one--dimensional loop. Then each plane is
decomposed into a sequence of lines, perpendicular to the $x_1$-- and
the $x_2$--axis, and the analogous procedure is applied to obtain the
field in $x_2$--direction. The charges which occur here are the line
averages, where however the plane averages have been subtracted (the
latter have already been taken into account via $E_1$). Finally, the
lines are decomposed into sites, and $E_3$ is determined from the
remaining charges where both line and plane averages have been
subtracted.

For field changes, elementary closed loops on the faces of the unit
cells (plaquettes) are considered. For the orthorombic lattice, these
are comprised of four nodes and respective links, such that each node
is connected to two plaquette links. Now, these four fields are
modified in such a way that the flux on each link is changed by the
same amount (taking into account the orientation along the closed
loop). Therefore, Gauss' law will still be satisfied after that move,
since at every node there will be some more flux entering but also the
same amount of flux leaving. Let us, for example, consider a plaquette
perpendicular to the $x_3$--axis, with a sequence of fields $E_1$,
$E_2'$, $E_1'$, $E_2$ along the loop, where $E_1$, $E_1'$ are positive
if the field points in positive $x_1$--direction (and analogous for
$E_2$, $E_2'$). Then the field updates are given by
\begin{eqnarray}
  E_1  & \to & E_1  + \delta E_1  , \\
  E_2' & \to & E_2' + \delta E_2' , \\
  E_1' & \to & E_1' + \delta E_1' , \\
  E_2  & \to & E_2  + \delta E_2  ,
\end{eqnarray}
or, in terms of fluxes,
\begin{eqnarray}
  \delta \phi_1  & = &  \Delta x_2 \Delta x_3 \delta E_1   = \delta \phi ,\\
  \delta \phi_2' & = &  \Delta x_1 \Delta x_3 \delta E_2'  = \delta \phi ,\\
  \delta \phi_1' & = &  \Delta x_2 \Delta x_3 \delta E_1'  = - \delta \phi ,\\
  \delta \phi_2  & = &  \Delta x_1 \Delta x_3 \delta E_2   = - \delta \phi ,
\end{eqnarray}
where the parameter $\delta \phi$ can be chosen arbitrarily without
violating Gauss' law. The associated change in the functional is given
by
\begin{eqnarray}
  \delta F \Delta V & = & \left\{ \left( \Delta x_1 \right)^2
                                + \left( \Delta x_2 \right)^2 \right\}
                          \left( \delta \phi \right)^2 \\
  \nonumber
                    & + & \Delta V \Delta x_1 \left( E_1 - E_1' \right)
                          \delta \phi \\
  \nonumber
                    & + & \Delta V \Delta x_2 \left( E_2 - E_2' \right)
                          \delta \phi ,
\end{eqnarray}
which is minimized for 
\begin{eqnarray}
  \delta \phi & = &
  \frac{1}{2} \frac{\Delta V}{
  \left( \Delta x_1 \right)^2 + \left( \Delta x_2 \right)^2} \\
  \nonumber
  & \times &
  \left\{ \Delta x_1 \left( E_1' - E_1 \right) -
          \Delta x_2 \left( E_2' - E_2 \right) \right\} .
\end{eqnarray}
This yields the optimal values for the field changes.

\begin{figure}
  \begin{center}
    \includegraphics[clip,width=0.47\textwidth]{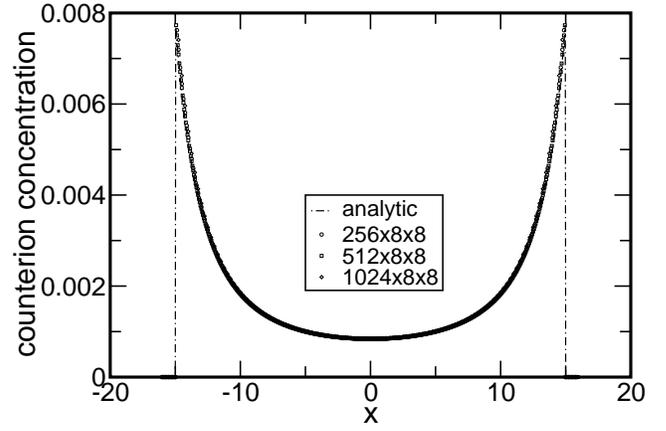}
  \end{center}
  \vspace{0.25cm}
  \begin{center}
    \includegraphics[clip,width=0.47\textwidth]{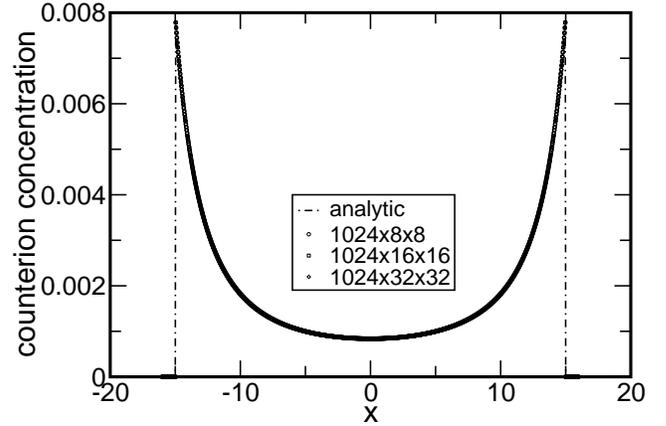}
  \end{center}
  \caption{Profile of the counterion concentration along the $x$--axis,
    for various grid resolutions as indicated by the legends.}
  \label{fig:concprofile_1d}
\end{figure}

For concentration moves between two adjacent nodes (connected by a
single link), Gauss' law is also conserved if the electric flux is
updated accordingly. Suppose that before the move the two adjacent
nodes $\mathbf{r}_0^{(A)}$ and $\mathbf{r}_0^{(B)}$ have,
respectively, concentrations $c^{(A)}$ and $c^{(B)}$ of some ionic
species with valence $z$. Without loss of generality we may assume
that node $B$ has a larger index value than node $A$. The electric
flux from node $A$ to node $B$ is then given by $(\Delta V / \Delta l)
E$, where $\Delta l$ is the length of the link, and $E$ the field on
it. Now, a certain (positive or negative) amount $\delta c$ is moved
from $A$ to $B$, i.~e.
\begin{eqnarray}
  c^{(A)} & \to & c^{(A)} - \delta c , \\
  c^{(B)} & \to & c^{(B)} + \delta c .
\end{eqnarray}
Gauss' law tells us that the flux should change by
$- \Delta V z \delta c$, and this is the case
for
\begin{equation}
  \delta E = - \Delta l \, z \, \delta c .
\end{equation}
The resulting change in the functional is given by
\begin{eqnarray}
  \frac{\Delta F}{\Delta V} & = & 
  \left(E + \frac{1}{2} \delta E \right) \, \delta E \\
  \nonumber
  & + & c^{(A)} \ln \left(1 - \frac{\delta c}{c^{(A)}} \right)
      + c^{(B)} \ln \left(1 + \frac{\delta c}{c^{(B)}} \right) \\
  \nonumber
  & - & \delta c \, \ln \frac{c^{(A)} - \delta c}{c^{(B)} + \delta c} .
\end{eqnarray}
In order to find the optimal value for $\delta c$, we minimize this
expression. The result is a nonlinear equation,
\begin{equation} 
  \label{eq:nonlinear}
  \delta c = \frac{c^{(A)} - c^{(B)}
  \exp \left(- z \Delta l (E - z \Delta l \delta c) \right) }
  {1 + 
  \exp \left(- z \Delta l (E - z \Delta l \delta c) \right) } ,
\end{equation}
which must be solved numerically. By introducing
\begin{eqnarray}
  c_+ & = & \frac{1}{2} \left( c^{(A)} + c^{(B)} \right), \\
  c_- & = & \frac{1}{2} \left( c^{(A)} - c^{(B)} \right), \\
  \xi & = & \frac{1}{2} z \Delta l \left( z \Delta l \delta c - E \right) ,
\end{eqnarray}
the equation is transformed to
\begin{equation}
  \tanh \xi
  + \frac{2}{ \left( z \Delta l \right)^2 c_+} \xi
  + \frac{E}{z \Delta l c_+} - \frac{c_-}{c_+} = 0 ,
\end{equation}
which shows that it has exactly one solution (the slope of the left
hand side is always positive). Since $-1 < \tanh \xi < 1$, 
the solution will satisfy the condition
\begin{equation}
  -1 <
  - \frac{2}{ \left( z \Delta l \right)^2 c_+} \xi
  - \frac{E}{z \Delta l c_+} + \frac{c_-}{c_+} < 1 ,
\end{equation}
which is equivalent to $c^{(A)} - \delta c > 0$, $c^{(B)} + \delta c >
0$, such that $\delta c$ will be in the physically admissible
range. Finally, the shape of the left hand side guarantees that a
Newton iteration starting at $\xi = 0$ will always converge; hence,
this rapid procedure was implemented.

\section{Numerical Results}
\label{sec:numerics}

In this section, the feasibility of the present approach is
demonstrated by two numerical examples. The choice of parameters is
inspired by previous computer simulations on electrokinetics done in
our group \cite{Lobaskin2004,Lobaskin2007,DuenwegetAl2008}. We
therefore quote them here in unscaled ``physical'' units, where
$\lambda_0$ denotes our elementary length scale (the Lennard--Jones
diameter in our simulations). All calculations are done with a Bjerrum
length $l_B = 1.3 \lambda_0$. The fixed charge distribution (boundary
condition) consists of positive charges only, while there is only one
ionic species, the monovalent counterions with valence $z = -1$.  
The resulting data are also given in ``physical'' units.

\begin{figure}
  \noindent
  \begin{centering}
      \includegraphics[clip,width=0.47\textwidth]{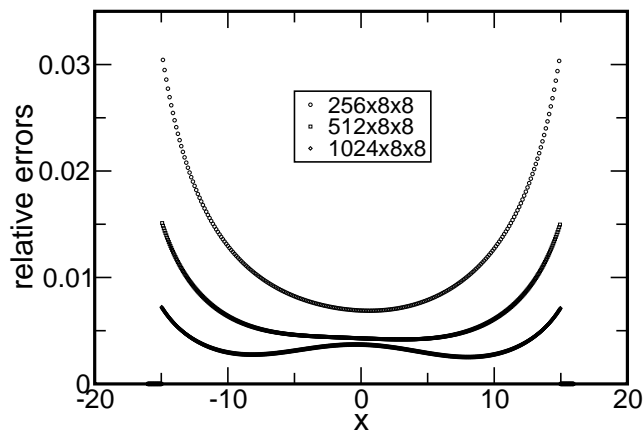}
  \end{centering}
  \caption{Relative errors of the concentration profiles, for different
           grid resolutions as indicated.}
  \label{fig:concerrors_1d}
\end{figure}

\subsection{Double Plane with Counterions}

Consider that the fixed charges are distributed in two infinite
parallel plates, perpendicular to the $x$-axis, placed at $x = -a$ and
$x = a$. The surface charge density in each plate is $\sigma$.
Furthermore, suppose that there are no salt ions and that the
counterions (of valence $z$, concentration $c(\mathbf{r})$) are
distributed in the region between the planes. Charge neutrality is
given by
\begin{equation}
  \int_{S} \sigma d \mathbf{s} + \int_{V} \, z\,
   c(\mathbf{r}) dV = 0.
\end{equation}

For this case, the problem is one--dimensional and its analytical
solution is well--known \cite{Engstrom1978}. Therefore, it is ideally
suited to test the algorithm. In reduced units, the Poisson--Boltzmann
equation is
\begin{eqnarray}
  \frac{d^{2} \psi}{dx^{2}} 
  & = & - A z \exp \left( - z \psi \right) \\
  \nonumber
  & = & - \frac{d}{d \psi} \left(
        - A \exp \left( - z \psi \right) \right) ,
\end{eqnarray}
which can be interpreted as Newton's equation of motion of a particle
with unit mass, whose coordinate is $\psi$ and where the time
corresponds to $x$. Hence, this equation can be solved via standard
methods of classical mechanics \cite{Landau}. The result is given by
\begin{eqnarray}
  \psi & = & \frac{2}{z} \ln \cos \left( s \frac{x}{a} \right) , \\
  c(x) & = & \frac{2}{z^2} \frac{s^{2}}{a^{2}} \cos^{-2}
             \left( s \frac{x}{a} \right) ,
\end{eqnarray}
where the parameter $s$ is related to $A$ via $2 s^2 = A z^2 a^2$.
The surface charge density is given by
\begin{equation}
  \sigma = \left. \frac{d \psi}{dx} \right\vert_{x=a} = 
  - \frac{2}{z} \frac{s}{a} \tan s ;
\end{equation}
therefore, $s$ can be obtained by solving a simple nonlinear equation
numerically.

\begin{figure}
  \noindent
  \begin{centering}
    \begin{tabular}{ccc}
      \includegraphics[clip,scale=0.23]{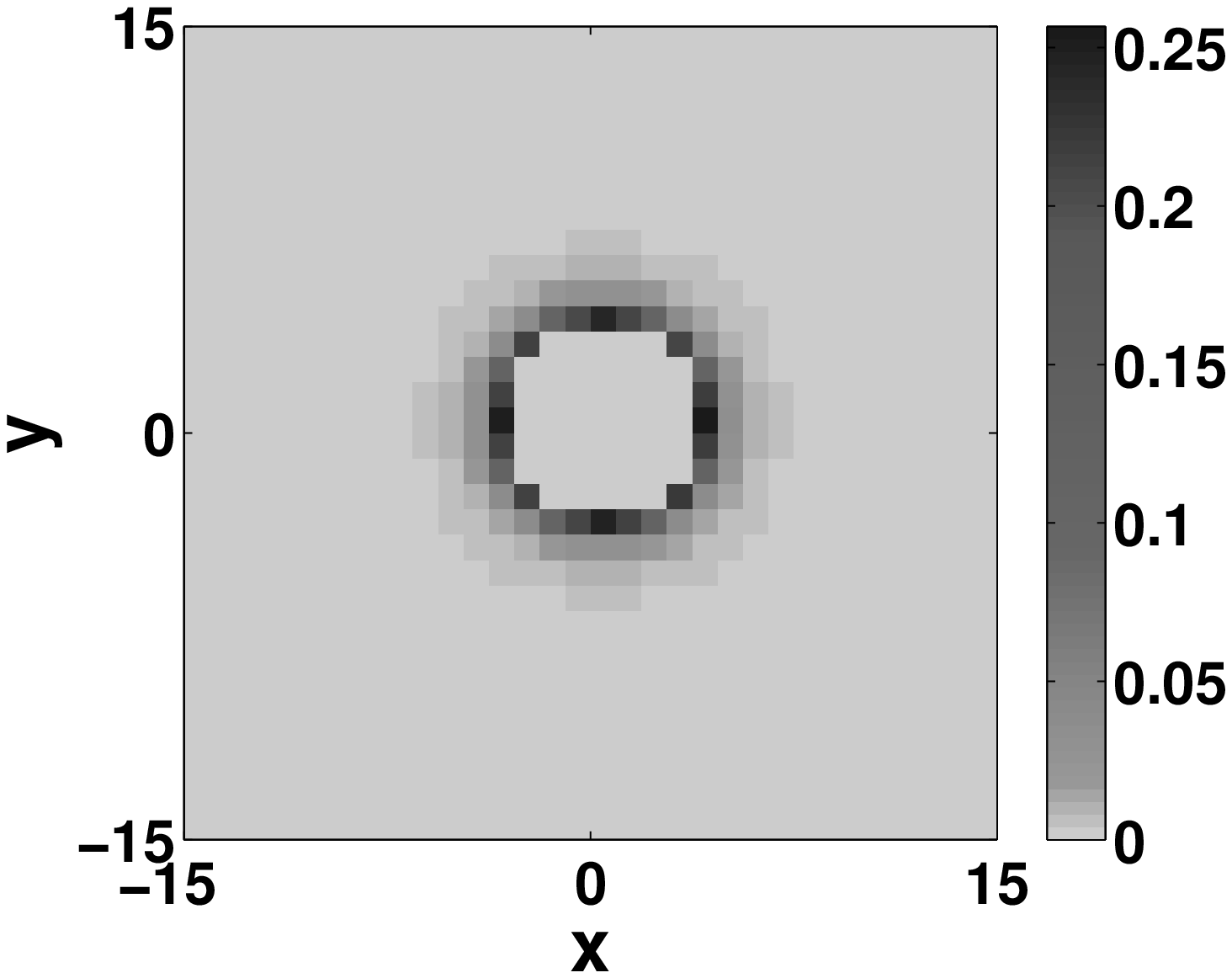} 
      & \hspace{0.1cm} & 
      \includegraphics[clip,scale=0.23]{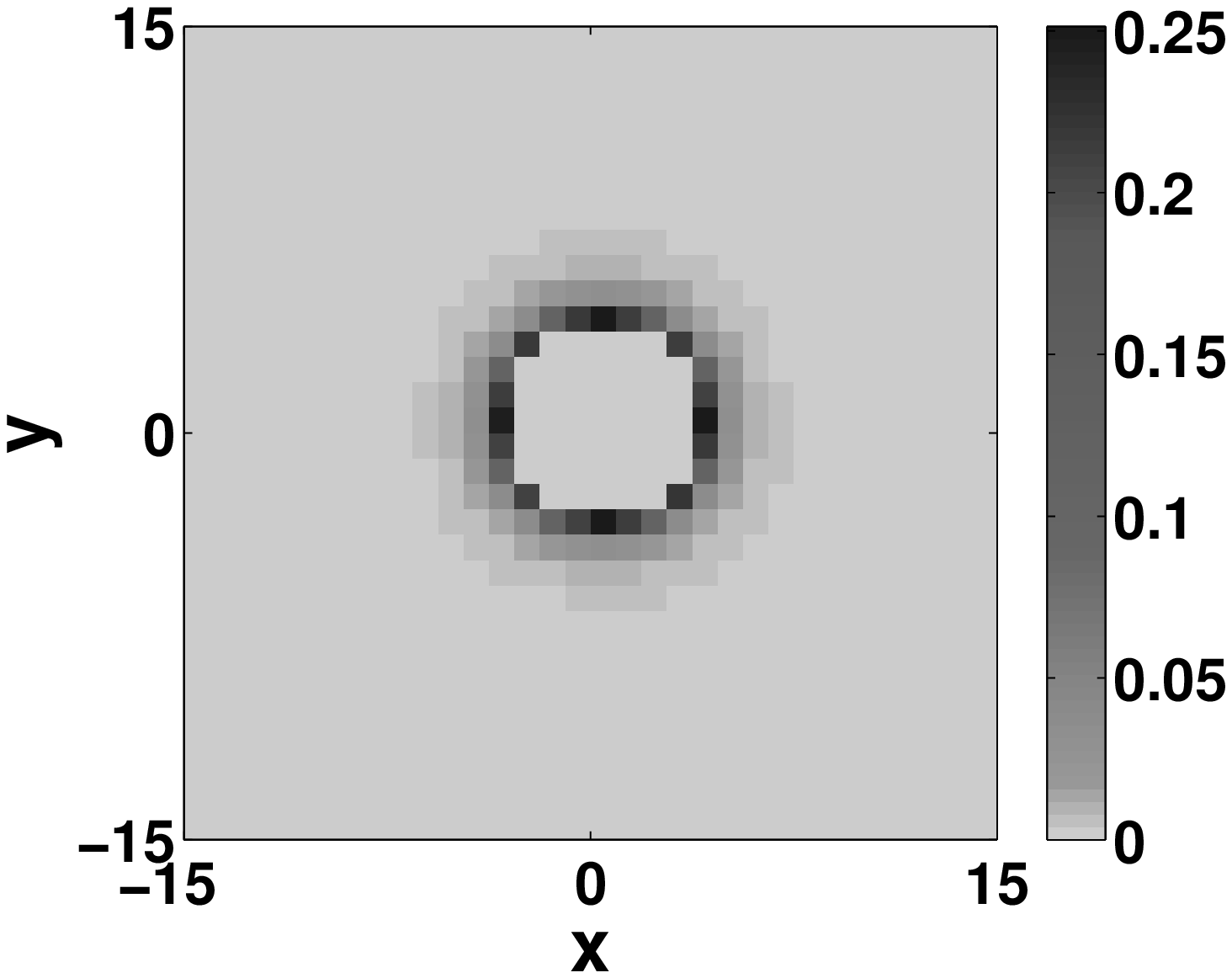}
      \tabularnewline
      Code A: $32 \times 32 \times 32$ 
      & \hspace{0.1cm} & 
      Code B: $32 \times 32 \times 32$
      \tabularnewline
      \includegraphics[clip,scale=0.23]{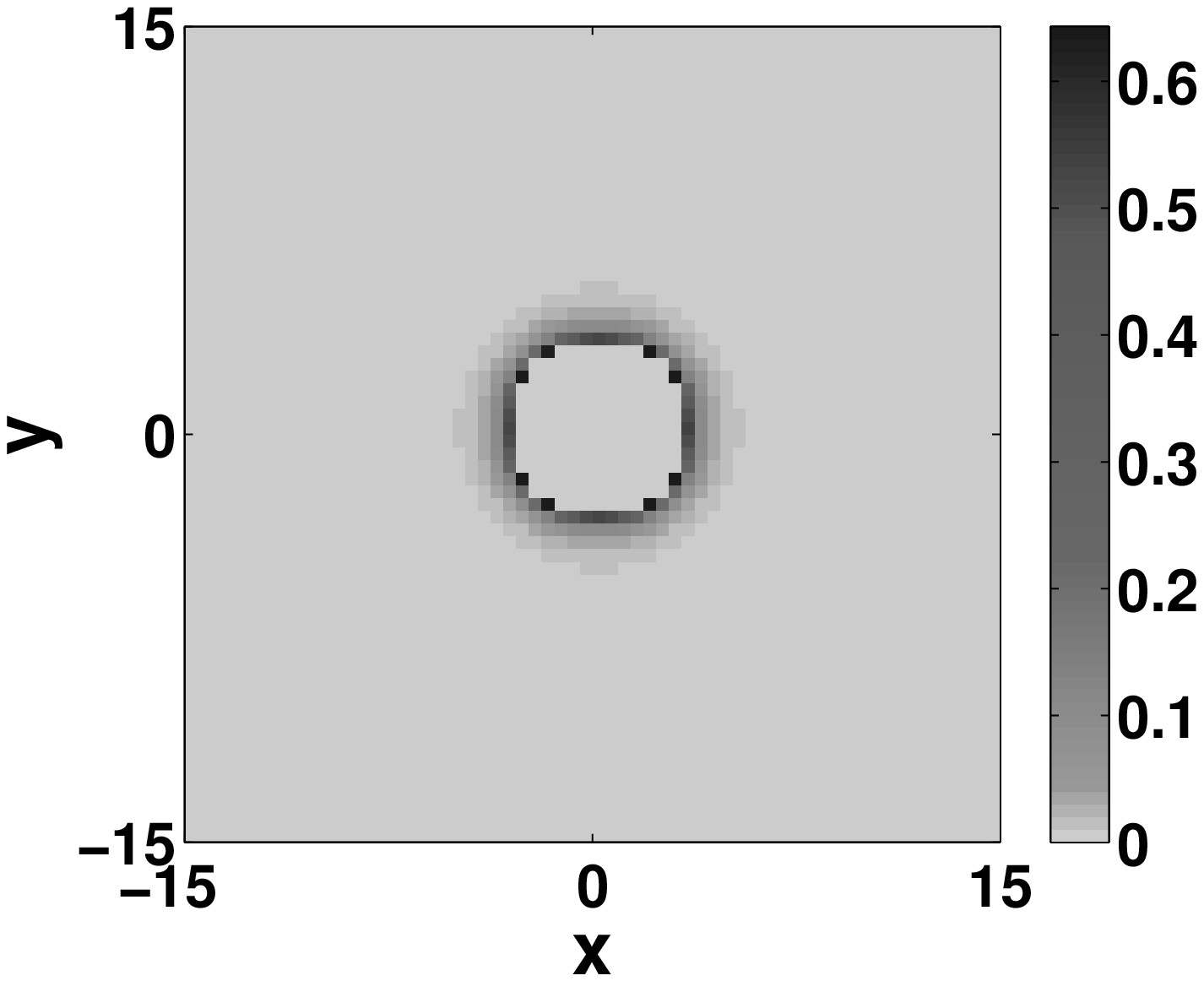}
      & \hspace{0.1cm} & 
      \includegraphics[clip,scale=0.23]{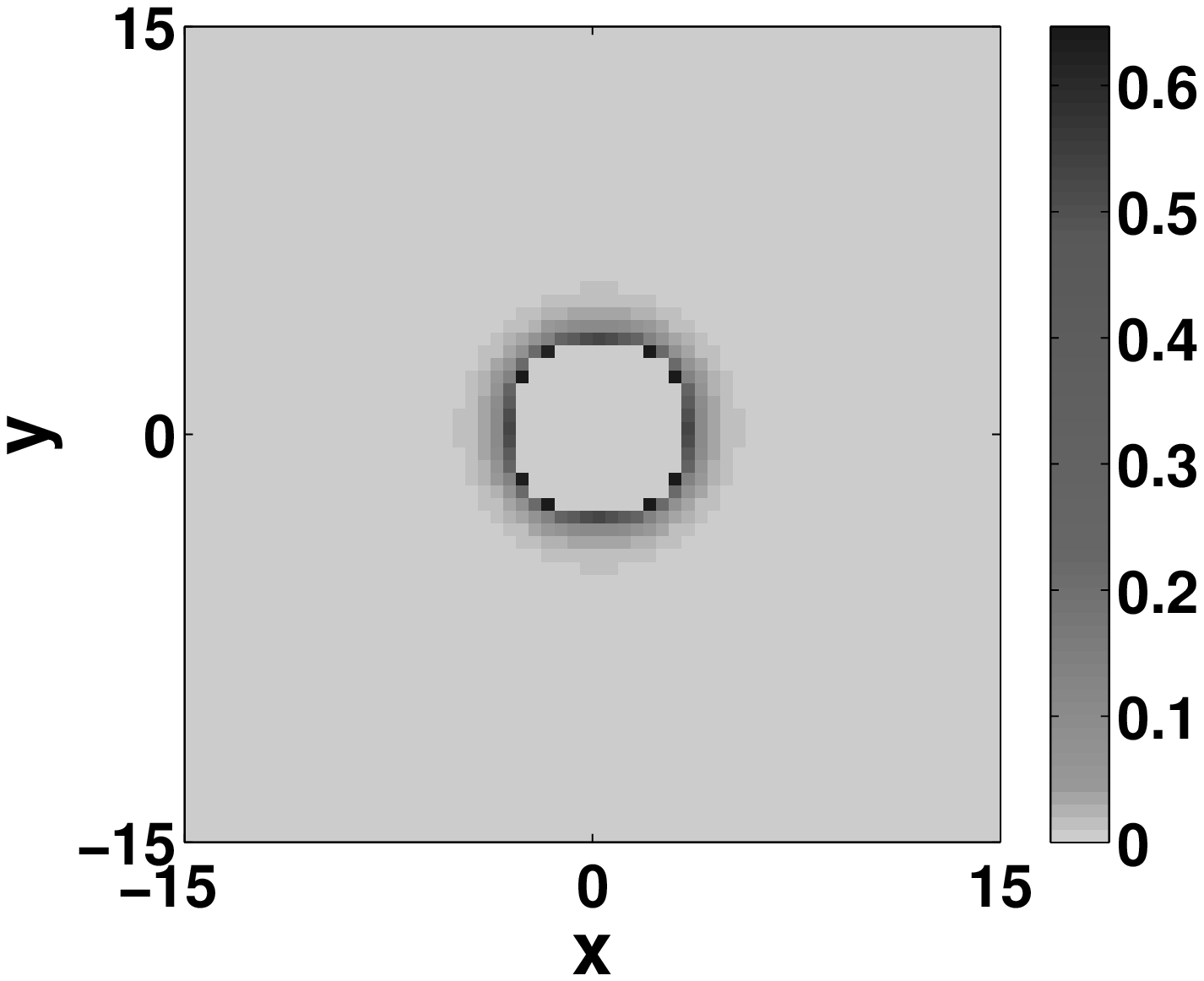}
      \tabularnewline
      Code A: $64 \times 64 \times 64$
      & \hspace{0.1cm} & 
      Code B: $64 \times 64 \times 64$
      \tabularnewline
      \includegraphics[clip,scale=0.23]{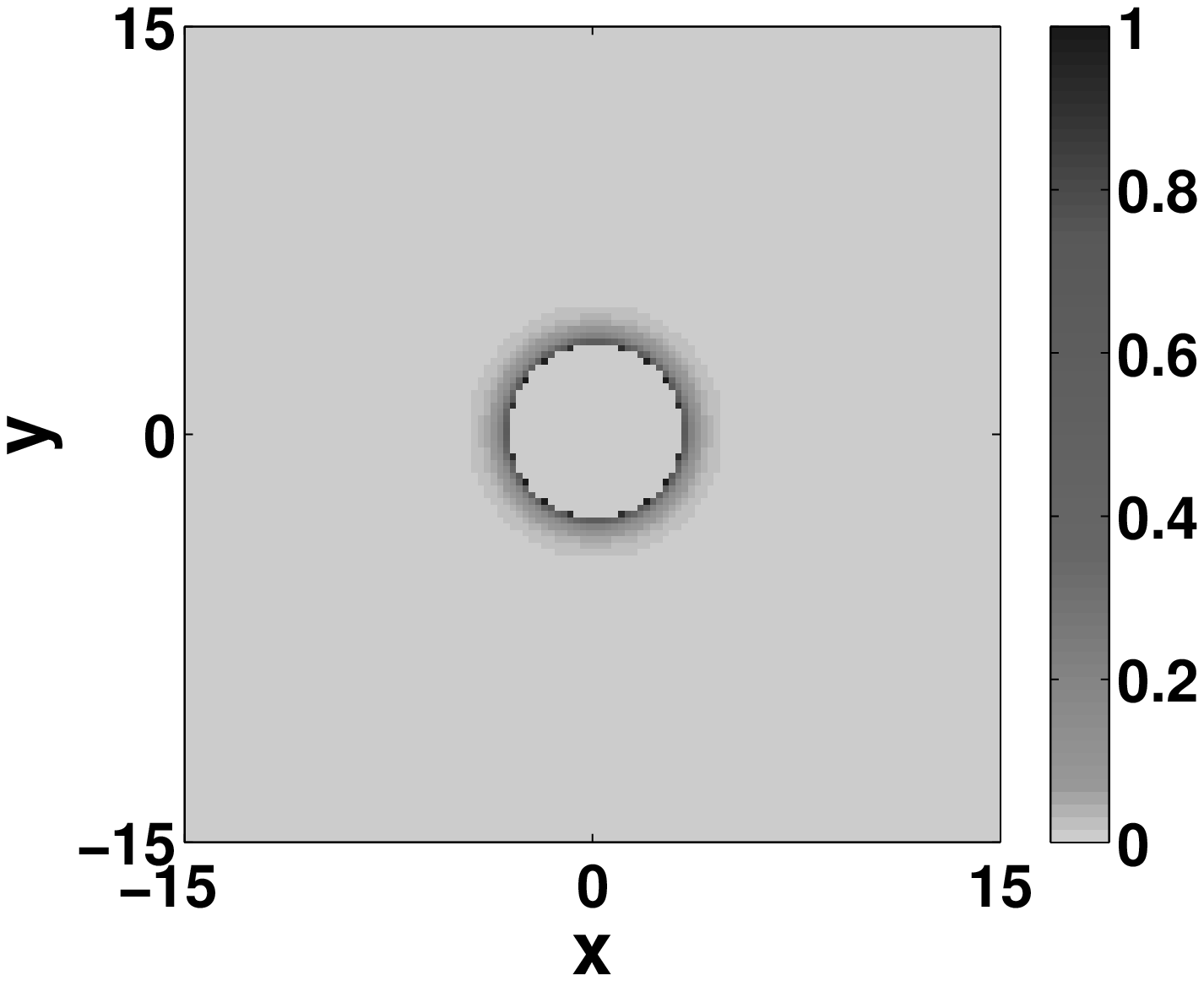}
      & \hspace{0.1cm} & 
      \includegraphics[clip,scale=0.23]{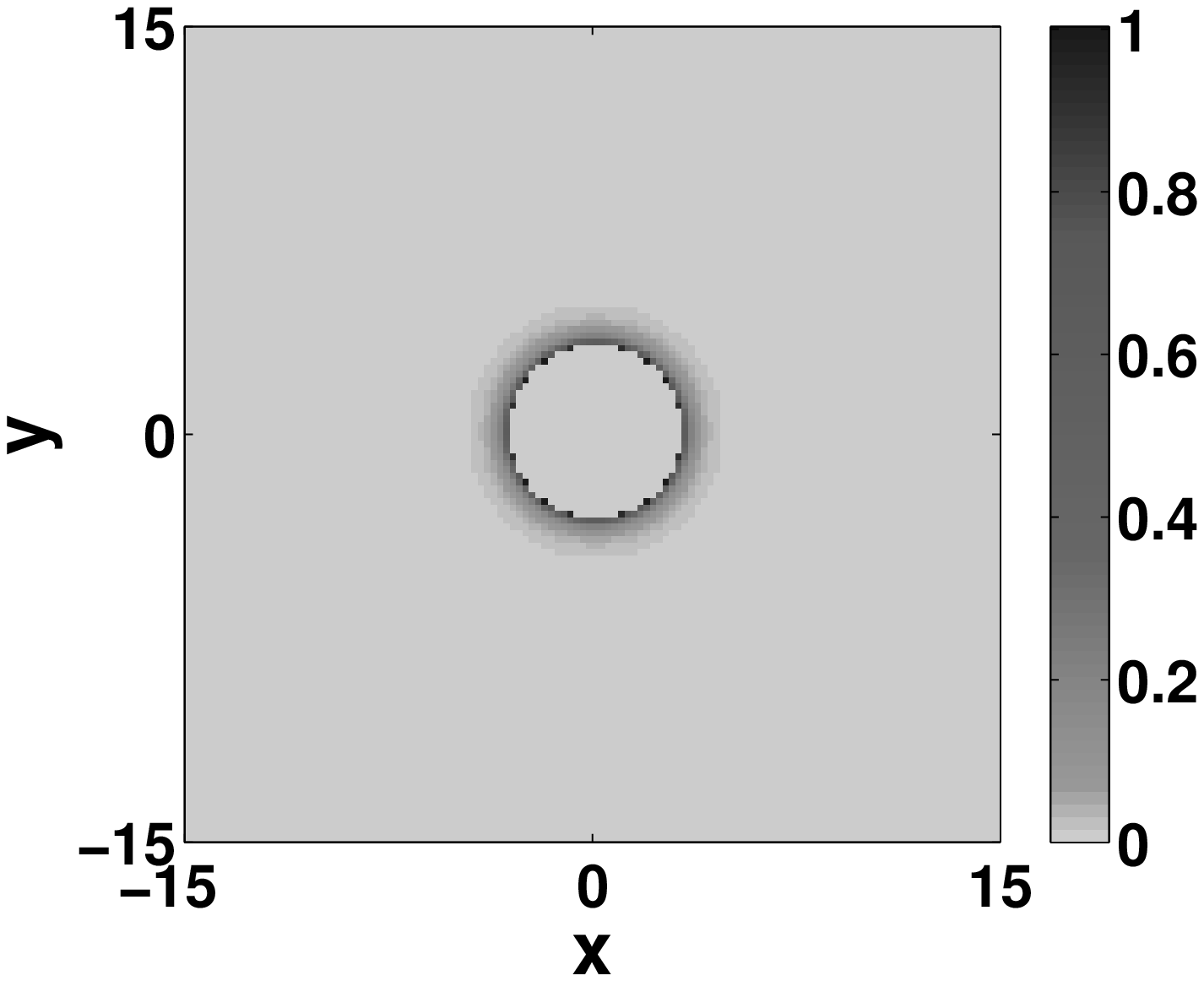}
      \tabularnewline
      Code A: $128 \times 128 \times 128$
      & \hspace{0.1cm} & 
      Code B: $128 \times 128 \times 128$
      \tabularnewline
    \end{tabular}
    \par
  \end{centering}
  \caption{Counterion concentration for a charged colloid.}
  \label{fig:Counterion-concentration-for}
\end{figure}

The algorithm presented can be used to solve this one--dimensional
problem. The planes are placed in the periodic box of size $l_1 \times
l_2 \times l_3$, at $x_1 = \pm a$. The amount of fixed charges, $Z e$,
is distributed homogeneously in the planes, so that the surface charge
density is $\sigma = Z e / (2 l_{2} l_{3})$. The corresponding
counterions are first distributed homogeneously in the region between
the planes. The simulation box in $x_1$ direction is somewhat larger
than $2 a$, in order to be able to implement periodic boundary
conditions in this direction. Beyond the charged planes, both the
concentration and the electric field vanish. Due to the periodic
boundary conditions in $x_2$ and $x_3$ direction, the system is
translationally invariant in these directions, and hence the solution
is one--dimensional.

The calculations were performed for $l_1 = l_2 = l_3 = 32 \lambda_0$,
$a = 15 \lambda_0$, $Z = 60$, and different grid sizes. The agreement
between the simulation and the analytic expression is quite good, see
Fig.~\ref{fig:concprofile_1d}, and increases with the grid resolution
in $x$ direction. Increasing the grid resolution in the orthogonal
directions has no effect, as expected (see
Fig.~\ref{fig:concprofile_1d}, lower part).
Figure~\ref{fig:concerrors_1d} shows the relative error for different
resolutions, defined by
\begin{equation}
    \mathrm{relative \, error} \, = \,
    \left| \frac{c_a (x) - c_{num} (x)}{c_a (x)} \right|  ,
\end{equation}
where $c_a (x)$ denotes the counterion concentration from the analytic
solution and $c_{num} (x)$ is the numerical result.

\begin{figure}
  \noindent
  \begin{centering}
      \includegraphics[clip,width=0.47\textwidth]
       {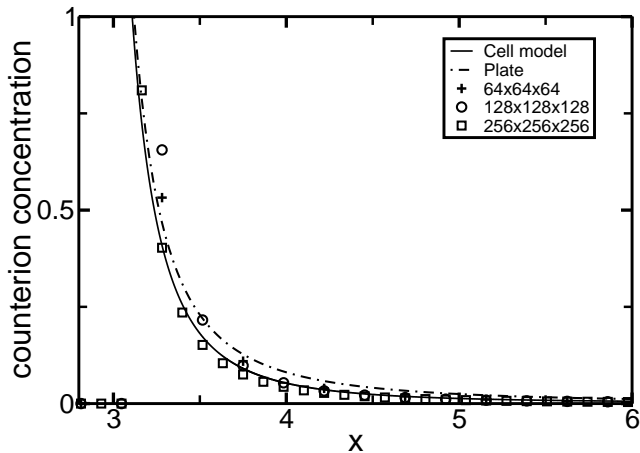}
    \par
  \end{centering}
  \caption{Counterion concentration profile along the $x$--axis
           for various grid resolutions as indicated by the legend.
           The continuous line is the solution of the $1$d isotropic
           cell model. The dash--dotted line shows the concentration
           profile for a $2$d charged plate with the same surface
           charge density.}
  \label{fig:concprofile_colloid}
\end{figure}

\subsection{Colloidal Particle in a Box}

In this case a colloidal particle, modeled by a sphere of radius $r$,
is placed at the center of the periodic box of size $l_{1} \times
l_{2} \times l_{3}$. The sphere carries a total charge of $Z e$,
uniformly distributed over its surface. In the simulation, the fixed
charges must be interpolated onto the nodes of the grid. The simplest
way is to generate $M \gg 1$ random points distributed uniformly over
the sphere surface. For each point, a (volume) charge density of $Z e
/( M \Delta V)$ is added to the closest grid node. These nodes are
marked as fixed.

The calculations were performed for a cubic box of sides 
$l_{1} = l_{2} = l_{3} = 30 \lambda_0$. Grids with cubic symmetry 
and various resolutions were used. A colloidal sphere of radius 
$r = 3 \lambda_0$ and valence $Z = 60$ was placed at the center of the box,
and the counterions were initially distributed uniformly over the outer space.

Two independent versions of the algorithm were implemented, one in C++
\cite{ekeproject} and another one in C. Runs for grids of different
sizes were done on an Intel Core 2 Duo E6600 FSB 1066 2x2.4 Ghz, with
4GB RAM. The algorithm was run until the change in the functional
reached a value smaller than $10^{-8}$. Performance results are
summarized in Table \ref{tab:Comparison}.

\begin{figure}
  \noindent
  \begin{centering}
      \includegraphics[clip,width=0.47\textwidth]
       {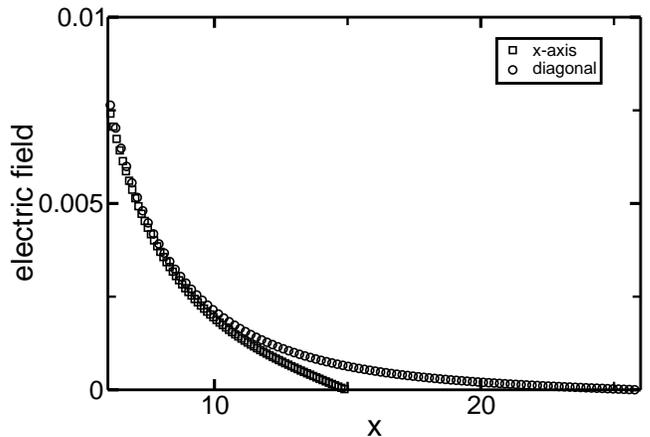}
    \par
  \end{centering}
  \caption{Electric field $E$ along the $x$--axis and along
           the $(111)$ diagonal for a grid resolution of
           $256^3$.}
  \label{fig:Eprofile_colloid}
\end{figure}

The pictures in Fig. \ref{fig:Counterion-concentration-for} show a
two--dimensional cut of the counterion concentration in a plane
perpendicular to the $z$--axis, while in
Fig. \ref{fig:concprofile_colloid} one--dimensional cuts (along the
$x$--direction of the simulation box) are shown. The data show that
near the surface of the charged colloid the discretization effects due
to the cubic grid are quite large; nevertheless the data for the
$256^3$ grid seem reasonably well converged to the continuum
limit. The concentration profiles in (100) direction and in (111)
direction are very close (i.~e. within the resolution of the plot the
curves coincide --- these data have however not been included for the
sake of clarity of the plot). Therefore one should expect that the
solution is essentially identical to one obtained with strict
spherical symmetry. This is indeed the case, as a comparison of the
concentration profile with the corresponding solution of the
spherically symmetric Poisson--Boltzmann cell model shows (see
Fig. \ref{fig:concprofile_colloid}).

\begin{table*}
  \noindent 
  \begin{centering}
    \begin{tabular}{|c||c|c||c|c|}
      \hline 
      & Code A (C++) & Code A (C++) & Code B (C) & Code B (C)
      \tabularnewline
      \hline 
      Grid Resolution & Time (s) & Memory (\%)  & Time (s) & Memory (\%) 
      \tabularnewline
      \hline
      \hline 
      $32\times32\times32$ & $30$ & $0.2$ &  $30$ & $0.1$ 
      \tabularnewline
      \hline 
      $64\times64\times64$ & $836$ & $0.5$ &  $844$ & $0.5$ 
      \tabularnewline
      \hline 
      $128\times128\times128$ & $20701$ & $2.9$ & $20574$ & $3.6$ 
      \tabularnewline
      \hline
      $256\times256\times256$ & $381371$ &  $22.5$ & $389030$ & $24.4$ 
      \tabularnewline
      \hline
      \hline
      & Iterations & Functional & Iterations & Functional
      \tabularnewline
      \hline
      \hline 
      $32\times32\times32$ & $642$ & $1114.83$ & $634$ & $1113.88$ 
      \tabularnewline
      \hline 
      $64\times64\times64$ & $2265$ & $1024.14$ & $2226$ & $1024.19$ 
      \tabularnewline
      \hline 
      $128\times128\times128$ & $7206$ & $957.63$ & $7038$ & $957.63$  
      \tabularnewline
      \hline
      $256\times256\times256$ & $20323$ & $923.85$ & $19711$ & $924.70$ 
      \tabularnewline
      \hline
    \end{tabular}
    \par
  \end{centering}
  \caption{Performance data for our two basic implementations.}
  \label{tab:Comparison}
\end{table*}

In the latter, the cubic simulation cell is replaced by a spherical
cell of the same volume, and the Poisson--Boltzmann equation is solved
for the radial coordinate. In our calculations, this was done by
transforming Eq. \ref{eq:compact_poisson_boltzmann} to spherical
coordinates, and then to a set of two coupled first--order
differential equations (one for the potential, one for the
field). This set was solved by a simple integrator analogous to the
velocity Verlet scheme known from Molecular Dynamics \cite{alltild},
using a step size of $\Delta r = 10^{-4}$ (in reduced units), and
integrating from the colloid radius outwards. The constant $A$
appearing in Eq. \ref{eq:compact_poisson_boltzmann} was determined
self--consistently by a shooting procedure, using the requirement that
the electric field must vanish at the outer radius, as a result of
Gauss' law and the overall charge neutrality. The thus--determined
profile agrees quite well with the one obtained from our algorithm
for the finest resolution.

Nevertheless, the solution for the cubic geometry does exhibit some
anisotropy that, by construction, is absent in the cell model. This
is essentially invisible in the concentration profile, but clearly
observable in the electric field profile, as shown in Fig.
\ref{fig:Eprofile_colloid}, where the decays in (100) and (111)
direction are compared.

In the immediate vicinity of the colloidal surface, one may view the
geometry as effectively planar. For a planar surface, the solution is
characterized by the so--called Gouy--Chapman length
\cite{grosberg_overcharging}. In our reduced units, this length has
the value $0.044$, which should be compared to the colloid radius
($0.5728$), and the lattice spacing ($0.045$ for the $128^3$
grid). The planar solution is shown in Fig.
\ref{fig:concprofile_colloid} (dash--dotted line); it also agrees
reasonably well with the profiles from our algorithm. Altogether, the
data indicate that a lattice spacing of roughly half a Gouy--Chapman
length is small enough to yield a reasonably well converged solution.

\section{Speedups}
\label{sec:speedup}

\begin{table*}
  \noindent
  \begin{centering}
    \begin{tabular}{|c||c|c||c|c|}
      \hline
      & Pure FFT & Pure FFT & FFT $+$ pre--conditioner 
      & FFT $+$ pre--conditioner 
      \tabularnewline
      \hline
      Grid Resolution & Time (s) & Number of iterations 
                      & Time (s) & Number of iterations 
      \tabularnewline
      \hline
      \hline
      $32\times32\times32$ & $14$ & $297$ &  $11$ & $242$
      \tabularnewline
      \hline
      $64\times64\times64$ & $409$ & $1150$ &  $331$ & $888$
      \tabularnewline
      \hline
      $128\times128\times128$ & $12180$ & $4427$ & $9744$ & $3421$
      \tabularnewline
      \hline
      $256\times256\times256$ & $354783$ & $16766$ & $273538$ & $12430$
      \tabularnewline
      \hline
    \end{tabular}
    \par
  \end{centering}
  \caption{Performance data for our two speeded--up implementations.
           Note that the data for the runs with pre--conditioner
           mean (i) CPU time for the overall procedure, and
           (ii) number of iterations in the final run with
           the finest resolution.} 
  \label{tab:speedup}
\end{table*}

As one sees from Tab. \ref{tab:Comparison}, the number of necessary
iterations and the amount of CPU time are quite large. We have
therefore looked for strategies to speed up the procedure without
sacrificing the basic formulation that provides intrinsic stability.
One possibility is to remove the rotational component of the electric
field not by means of plaquette moves but rather by solving the
Poisson equation. Within the chosen discretization scheme, the
electrostatic potential needs to be an object associated with the
sites, such that the electric field on a link is obtained by simply
taking the potential difference between the adjacent sites. This leads
to the simplest possible finite--difference scheme for the Poisson
equation, which can be solved efficiently and in a stable way by using
a Fast Fourier Transform (FFT) and the appropriate lattice Green's
function \cite{numrecip}. This has the advantage that one single
lattice sweep not only reduces the rotational component (as is
the case for the plaquette moves) but rather eliminates it
completely. Therefore the FFT promises to increase the
convergence speed. An easy implementation is possible using
the well--known and efficient fftw3 library routine \cite{fftw3}. It
should be noted that the link moves, which update the concentrations
and the fields simultaneously, remain unchanged, such that the
procedure still stays strictly on the constraint surface.

In a first implementation, we eliminated all plaquette moves 
and replaced them with FFT sweeps done during initialization as well as 
subsequently after every 25th link sweep. As seen from Tab. \ref{tab:speedup},
this improves the efficiency roughly by a factor of $1.1 \dots 2$.
These results were obtained on the same computer as those of
Tab. \ref{tab:Comparison}.

Furthermore, we can tackle the slowdown that comes from the fact that
the ions have to be moved by site--by--site hops throughout the system
(``hydrodynamic slowing down''). To this end, we first run the
calculation on a rather coarse grid (in practice, we started with $8
\times 8 \times 8$), such that most of the necessary ``mass
transport'' is already done in that preliminary run. Starting from
there, we go to a finer grid (in practice, we reduced the lattice
spacing in all three directions by a factor of two) and linearly
interpolate the output of the previous run onto that grid. Then the
free energy is relaxed again; the output of that run is interpolated
onto a yet finer grid, and so on.  Obviously, this can be done rather
easily by a straightforward recursion, until the desired grid
resolution is reached. The runs before the finest resolution may then
be viewed as a ``pre--conditioner''. This optimization yields another
speedup by roughly 25$\%$, as seen from Tab. \ref{tab:speedup}.

These are obviously two rather simple optimizations, which do not
interfere with the basic data structure of the simple Cartesian
grid. Further optimizations, which are however much more complicated,
are possible by (i) adaptive mesh refinement (i.~e. a fine resolution
is only used in those regions where the fields vary strongly), and
(ii) using finite--element--type unstructured grids. Such more
advanced approaches would be based upon constructing the dual
(Voronoi) lattice in order to define and calculate the fluxes. While
this is expected to yield further substantial speedups, this was not
attempted here, and is rather mentioned as a suggestion for the larger
community.

\section{Concluding Remarks}
\label{sec:conclus}

Variational techniques were used to develop an algorithm for the
Poisson--Boltzmann equation. The required amount of memory scales
linearly with the system size. In our first simple implementation, all
moves are local. This leads to fast memory access and fast
calculations at each move. Furthermore, the algorithm can be easily
parallelized. On the other hand, the charges need time to move between
distant nodes. In fact, while the time spent on each move is
essentially constant, the total number of sweeps required depends
highly on how large a grid is swept. Considerable speedups were
possible by using FFTs and a hierarchical pre--conditioner, but the
basic ``hydrodynamic slowing down'' remains still present. Further
speedups are likely to be possible by adaptive mesh refinement and by
using unstructured grids.

Existing algorithms based on standard discretizations of the
differential operators are probably significantly faster than even the
fastest of our current implementations. Nevertheless, the approach
that we have outlined in the present paper has the inherent advantage
that its mathematical formulation provides \emph{intrinsic}
stability. This is mainly due to the fact that the free energy
functional is formulated in terms of the electric field instead of the
electrostatic potential, which provides a search for a \emph{true
  minimum} in function space, rather than for a saddle point. As a
result, the algorithm is very robust, in the sense that it will
\emph{always} converge to the solution, and at every iteration it
approaches the solution more closely. Furthermore, the essential
conservation laws that are at the heart of electrostatics ---
conservation of mass and conservation of electric flux --- are built
into the formulation with machine accuracy. We believe that these are
fundamental advantages directly related to the underlying physics of
the problem, and we consider them as much more important than
implementation details. The Maggs formulation provides a new way of
thinking about electrostatics, and we hope that this, in combination
with existing numerical ``technology'', will in the future bring about
very useful algorithms.

\acknowledgments{This work was funded by the SFB TR 6 of the
  Deutsche Forschungsgemeinschaft. Stimulating discussions with
  B. Li are gratefully acknowledged.}


\end{document}